\documentclass{article}

\PassOptionsToPackage{numbers, compress}{natbib}

\usepackage[final]{neurips_2019}

\usepackage[utf8]{inputenc} 
\usepackage[T1]{fontenc}    
\usepackage{hyperref}       
\usepackage{url}            
\usepackage{booktabs}       
\usepackage{amsfonts}       
\usepackage{nicefrac}       
\usepackage{microtype}      

\usepackage[pdftex]{graphicx}
\title{Transfer Learning in 4D for Breast Cancer Diagnosis using Dynamic Contrast-Enhanced Magnetic Resonance Imaging}

\author{%
  Qiyuan Hu, Heather M. Whitney, Maryellen L. Giger\\
  Committee on Medical Physics, Department of Radiology\\
  University of Chicago\\
  Chicago, IL 60637 \\
  \texttt{qhu@uchicago.edu} \\
}

\begin{document}

\maketitle

\begin{abstract}
  Deep transfer learning using dynamic contrast-enhanced magnetic resonance imaging (DCE-MRI) has shown strong predictive power in characterization of breast lesions. However, pretrained convolutional neural networks (CNNs) require 2D inputs, limiting the ability to exploit the rich 4D (volumetric and temporal) image information inherent in DCE-MRI that is clinically valuable for lesion assessment. Training 3D CNNs from scratch, a common method to utilize high-dimensional information in medical images, is computationally expensive and is not best suited for moderately sized healthcare datasets. Therefore, we propose a novel approach using transfer learning that incorporates the 4D information from DCE-MRI, where volumetric information is collapsed at feature level by max pooling along the projection perpendicular to the transverse slices and the temporal information is contained in either in second-post contrast subtraction images. Our methodology yielded an area under the receiver operating characteristic curve of \(0.89\pm0.01\) on a dataset of 1161 breast lesions, significantly outperforming a previously approach that incorporates the 4D information in DCE-MRI by the use of maximum intensity projection (MIP) images.
\end{abstract}

\section{Introduction}

Magnetic resonance imaging (MRI) is one of the imaging modalities for clinical diagnosis and monitoring of breast cancer. It has been established for use in screening patients with high risk of breast cancer, cancer staging, and monitoring cancer response to therapies \cite{pickles2015prognostic, turnbull2009dynamic}. In comparison with more commonly used clinical modalities, such as mammography and ultrasound, MRI offers much higher sensitivity to breast cancer diagnosis \cite{morrow2011mri, kuhl2005mammography}. Dynamic contrast-enhanced (DCE)-MRI provides high-resolution volumetric lesion visualization as well as functional information via temporal contrast enhancement patterns—information that carries significant clinical value for breast cancer management. However, the labor-intensive breast MRI interpretation process in combination with deficiency in MRI reading experts make it an expensive clinical procedure.

In order to assist radiologists in the interpretation of diagnostic imaging, automated computer-aided diagnosis (CADx) systems continue to be developed to potentially improve the accuracy and efficiency of breast cancer diagnosis \cite{giger2013breast}. Recently, deep learning methods have demonstrated success in computer-aided diagnostic and prognostic performance based on medical scans \cite{greenspan2016guest, tajbakhsh2016convolutional, shin2016deep}. Although training deep neural networks from scratch typically relies on massive datasets for training and is thus often intractable for medical research due to data scarcity, it has been shown that standard transfer learning techniques like fine-tuning or feature extraction based on ImageNet-trained convolutional neural networks (CNNs) can be used for CADx \cite{yosinski2014transferable, donahue2014decaf, huynh2016digital, antropova2017deep}. However, pretrained CNNs require two-dimensional (2D) inputs, limiting the ability to exploit high-dimensional volumetric and temporal image information that can contribute to lesion classification. 

To take advantage of the rich 4D information inherent in DCE-MRI without sacrificing the efficiency provided by transfer learning, a previously proposed method used the second post-contrast subtraction maximum intensity projection (MIP) images to classify breast lesions as benign or malignant, and showed superiority to using only 2D or 3D information \cite{antropova2018use}. In this study, we propose a new transfer learning method that also makes use of the 4D information in DCE-MRI, but instead of reducing the volumetric information to 2D at the image level via creating MIP images, we do so at the feature level by max pooling CNN features from all slices for a given lesion. Compared with using MIP images, our CNN feature MIP method demonstrated significant improvement in classification performance in the task of distinguishing between benign and malignant breast lesions. 

\section{Materials and Methods}
\subsection{Database}
A database consisting of 1161 unique breast lesions from 855 women who have undergone breast MR exams was retrospectively collected under HIPAA-compliant Institutional Review Board protocols. Of all lesions, 270 were benign (23\%) and 891 were malignant (77\%) based on pathology and radiology reports. Images in the database were acquired over the span of 12 years, from 2005 to 2017, using either 1.5 T or 3 T Philips Achieva scanners with T1-weighted spoiled gradient sequence.

\subsection{CNN Input}

\begin{figure}
  \centering
  \includegraphics[width=0.9\linewidth]{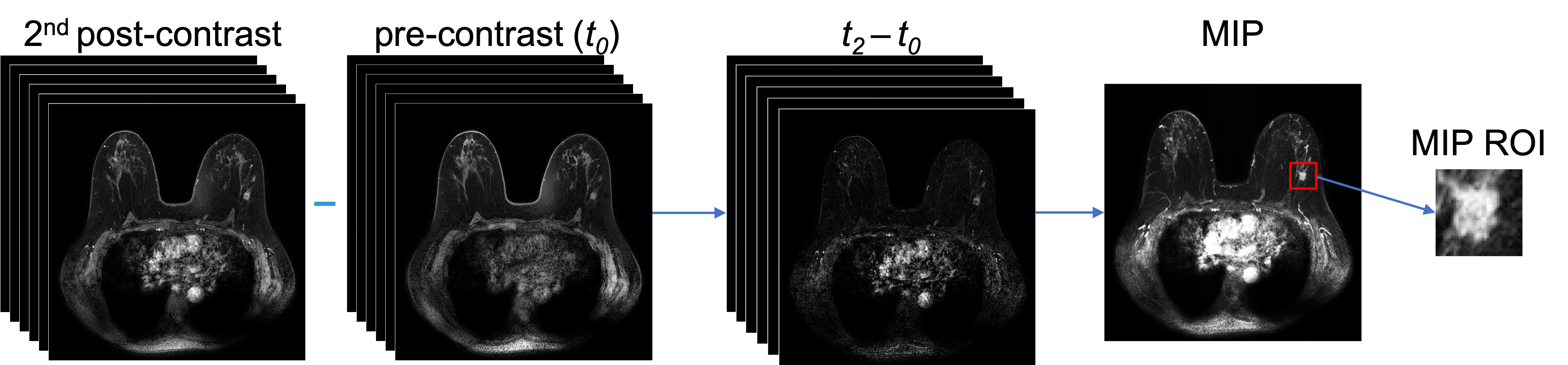}
  \caption{Illustration of the processes to construct the second post-contrast subtraction images, the subtraction maximum intensity projection (MIP) images, and region of interest (ROI).}
  \label{fig: MIPROI}
\end{figure}

As illustrated in Fig. \ref{fig: MIPROI}, the subtraction images were created by subtracting the pre-contrast (\(t_0\)) images from their corresponding second post-contrast (\(t_2\)) images in order to emphasize the contrast enhancement pattern within the lesion and suppress constant background. To generate MIP images, the 3D volume of subtracted images for each lesion was then collapsed into a 2D image by selecting the voxel with the maximum intensity along the axial dimension, i.e., perpendicular to the transverse slices. 

To avoid confounding contributions from distant voxels, a region of interest (ROI) around each lesion was automatically cropped from the image to use in the subsequent classification process. The ROI size was chosen based on the maximum dimension of each lesion and a small part of the parenchyma around the lesion was included. The minimum ROI size was set to \(32\times32\) pixels as required by the pre-trained CNN architecture. The cropping process is illustrated in Fig. \ref{fig: MIPROI}.

\subsection{Classification}
\begin{figure}
  \centering
  \includegraphics[width=0.4\linewidth]{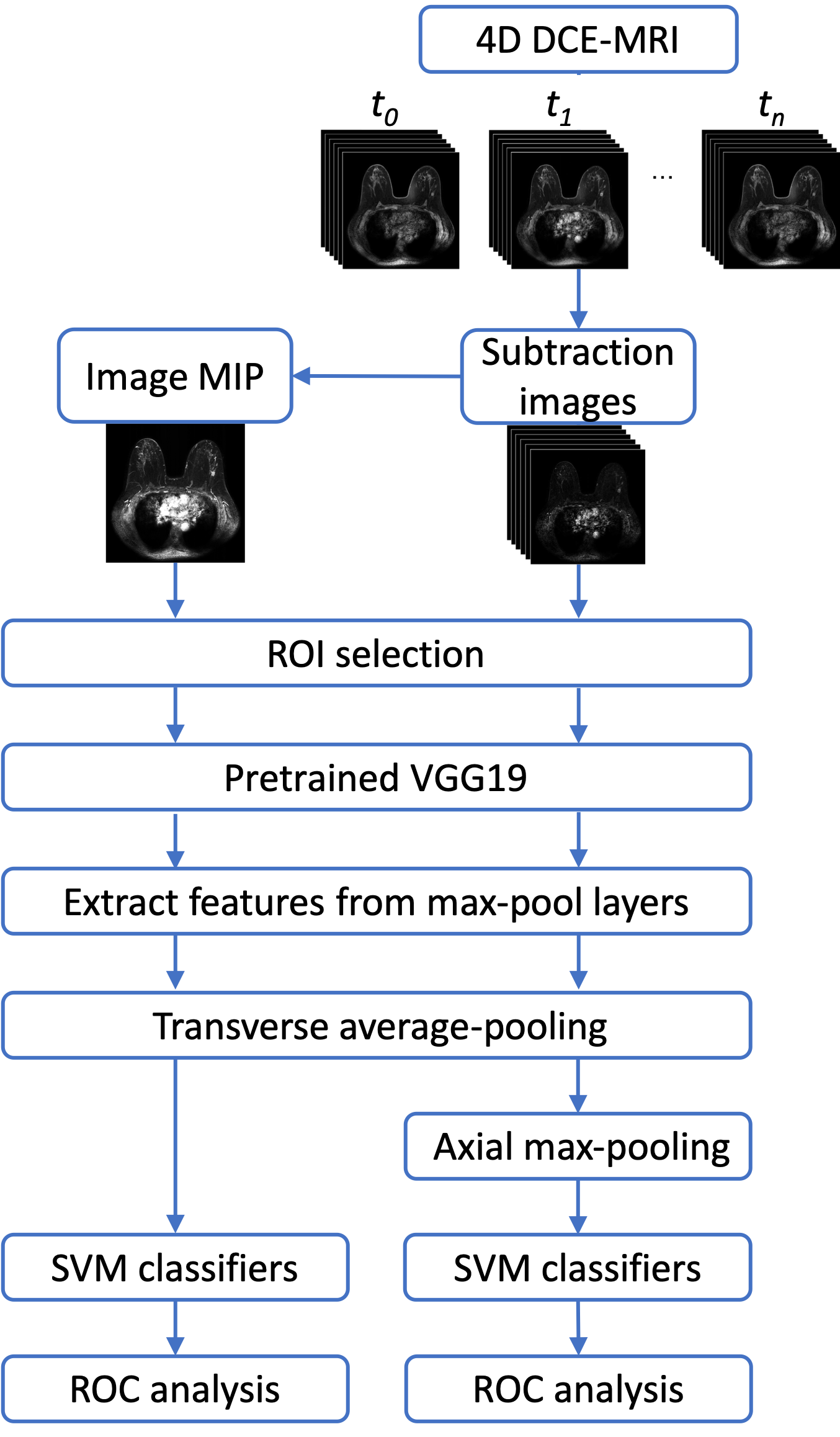}
  \caption{Lesion classification pipelines based on diagnostic images. Three-dimensional volumetric lesion information from dynamic contrast-enhanced (DCE)-MRI is collapsed into 2D at the image level by using maximum intensity projection (MIP) images (left), or at the feature level by max pooling CNN features extracted from all slices (right) along the axial dimension.}
  \label{fig: flowchart}
\end{figure}

Figure \ref{fig: flowchart} schematically shows the transfer learning classification and evaluation process for the two methods following the ROI construction. For each lesion, CNN features were extracted from both the MIP and all individual slices of the subtraction ROIs using a VGG19 model pretrained on ImageNet \cite{simonyan2014very, deng2009imagenet}. Feature vectors were extracted at various network depths from the five max pooling layers of the VGGNet. These features were then average-pooled along the spatial dimensions and normalized with Euclidian distance. The pooled features were then concatenated to form a CNN feature vector for a given lesion \cite{huynh2016digital, antropova2017deep}. For the CNN feature MIP method, features extracted from all slices were concatenated into a 3D feature vector and was then collapsed into a 2D feature vector by max pooling along the axial dimension, i.e., the direction perpendicular to the transverse slices. 

Two linear support vector machine (SVM) classifiers were trained on the CNN features extracted from subtraction MIPs and subtraction volumes separately to differentiate between benign and malignant lesions. SVM was chosen over other classification methods due to its ability to handle sparse high-dimensional data, which is an attribute of the CNN features \cite{russakovsky2015imagenet}. To address the problem of class imbalance, a misclassification penalty for cases in each class was assigned to be inversely proportional to its class prevalence in the training data. 

\subsection{Evaluation}
Each SVM classifier was trained and evaluated using nested five-fold cross-validation, where all lesions from the same patient were kept together in the same fold in order to eliminate the impact of using correlated lesions for both training and testing. Within each training set in the outer cross-validation, SVM hyperparameters were optimized by an internal grid search with an inner five-fold cross-validation \cite{Shawe-Taylor:2011:ROM:2304791.2305138}. Principal component analysis fit on the training set was applied to both training and test sets to reduce feature dimensionality \cite{jolliffe2011principal}. The cross-validation evaluation process was repeated 10 times with different random seeds, and the final prediction score for each lesion was averaged over the 10 repetitions. Class prevalence was held constant across the five cross-validation folds.

Classifier performances were evaluated using receiver operating characteristic (ROC) curve analysis, with area under the ROC curve (AUC) serving as the figure of merit \cite{Metz1998, metz1999proper}. The two classification methods were compared using the DeLong test \cite{delong1988comparing, sun2014fast}. Standard errors and 95\% confidence interval (CI) of the difference in AUCs were calculated by bootstrapping the posterior probabilities of malignancy \cite{efron1987better}.

\section{Results}
\begin{figure}
  \centering
  \includegraphics[width=0.7\linewidth]{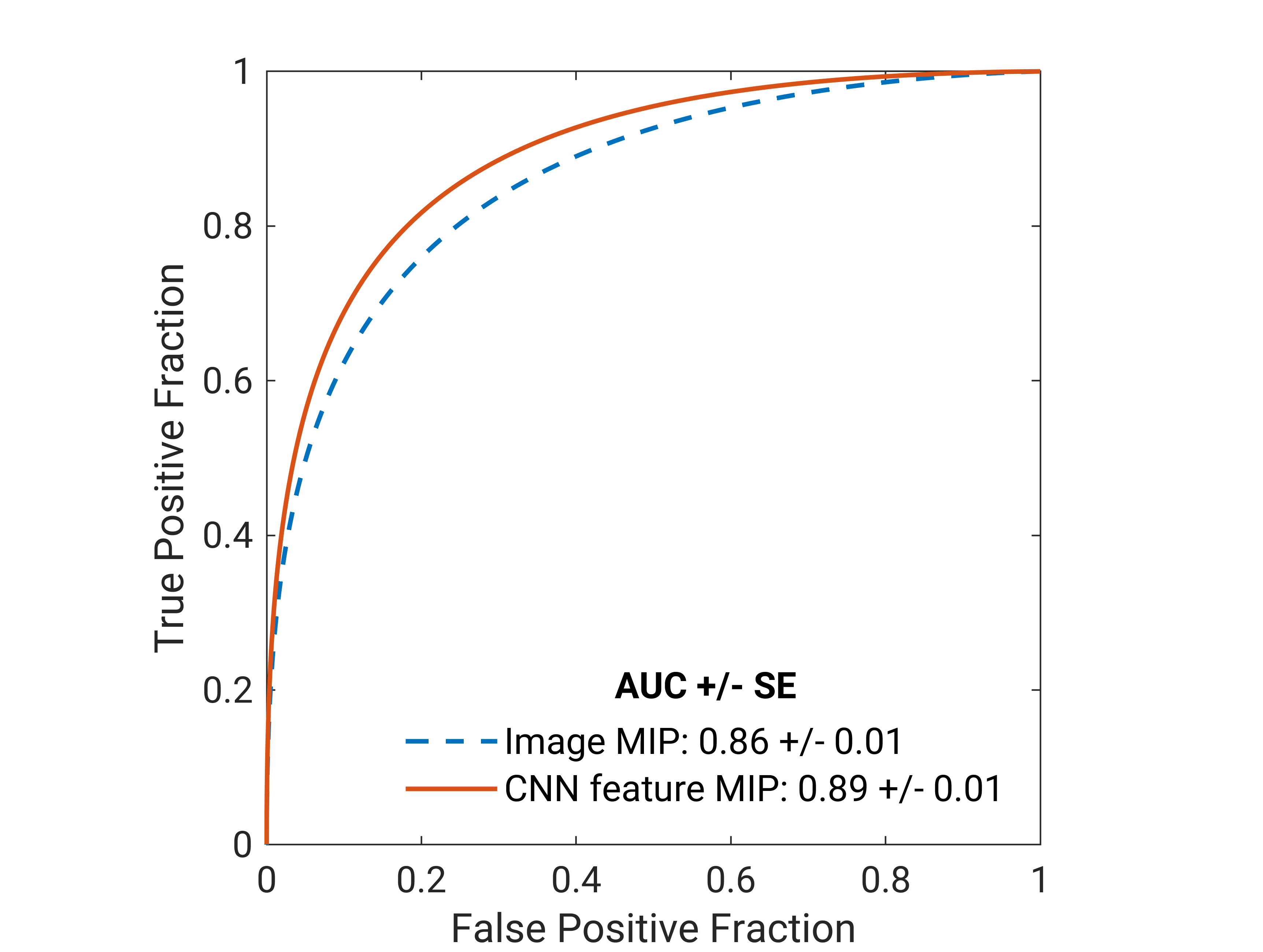}
  \caption{Fitted binomial receiver operating characteristic (ROC) curves for two classifiers that utilize the 4D information from dynamic contrast-enhanced (DCE)-MRI. The dashed blue line represents the image maximum intensity projection (MIP) method, where the 3D volumetric information is reduced to 2D at the image level. The solid orange line represents the CNN feature MIP method of collapsing the 3D volumetric information at the feature level by max pooling features from all slices. The legend gives the area under the ROC curve (AUC) with standard error (SE) for each classifier.}
  \label{fig: ROC}
\end{figure}

Figure \ref{fig: ROC} presents the ROC curves of the two classification methods in the task of distinguishing benign and malignant breast lesions. The 95\% CI of the difference in AUCs was $\Delta AUC = [0.01, 0.04]$ and the p-value was $P < .001$. The results suggest that in the task of distinguishing benign and malignant breast lesions using deep transfer learning, 3D volumetric information in DCE-MRI may have superior predictive power when collapsed along the axial dimension via MIP at the feature level than at the image level.

\section{Discussion}

In conclusion, our study proposed a novel methodology to incorporate the volumetric information inherent in DCE-MRI when using deep transfer learning. Our method of utilizing the volumetric information by max pooling features along the axial dimension significantly outperformed the previously proposed method of using MIP images and exhibited comparable computational efficiency.

High dimensionality and data scarcity are unique challenges in deep learning applications to medical imaging. In order to exploit the rich clinical information in medical images without sacrificing computational efficiency or model performance, it is important to devise approaches to use transfer learning in creative ways so that volumetric and temporal data can be incorporated even when networks pretrained on 2D images are used. Compared with recent studies that train 3D CNNs from scratch on DCE-MRI \cite{dalmis2019artificial,li2017discriminating}, our methodology of using transfer learning on 4D medical imaging data is computationally efficient, has demonstrated high performance on moderately sized datasets, and does not require intensive image preprocessing. Future work will further expand the analysis to include other valuable sequences in multiparametric MRI, rather than DCE-MRI alone. Furthermore, we would also like to increase the size of our database, which would allow us to explore fine-tuning and to use a standard training/validation/test split of the data. Finally, expanding the database to include images from other medical centers would allow us to perform validation on an independent, external dataset, and would help us develop a more robust system by including the heterogeneity resulting from different imaging manufacturers and facility protocols. 

\section*{Acknowledgements}
The authors acknowledge other lab members, including Karen Drukker, PhD, MBA; Alexandra Edwards, MA; Hui Li, PhD; and John Papaioannou, MS, Department of Radiology, The University of Chicago, Chicago, IL for their contributions to the datasets and discussions. The work was partially supported by NIH QIN Grant U01CA195564, NIH NCI R15 CA227948, and the RSNA/AAPM Graduate Fellowship. MLG is a stockholder in R2 technology/Hologic and QView, receives royalties from Hologic, GE Medical Systems, MEDIAN Technologies, Riverain Medical, Mitsubishi and Toshiba, and is a cofounder of and equity holder in Quantitative Insights (now Qlarity Imaging). It is the University of Chicago Conflict of Interest Policy that investigators disclose publicly actual or potential significant financial interest that would reasonably appear to be directly and significantly affected by the research activities.

\bibliography{myCitationLib}

\end{document}